\documentclass[aip,reprint]{revtex4-1}

\usepackage{graphicx}
\usepackage{setspace}
\usepackage{subfigure}
\usepackage{color,colortbl}
\usepackage{textcomp}
\usepackage{dcolumn}
\usepackage{bm}
\newcommand{\gst}{Ge$_{2}$Sb$_{2}$Te$_{5}$}
\newcommand{\gsta}{Ag$_{0.5}$Ge$_{2}$Sb$_{2}$Te$_{5}$}

\begin{document}

\title{Direct \textit{ab-initio} molecular dynamic study of ultrafast phase change in Ag-alloyed \gst}

\author{B. Prasai}
\author{G. Chen}
\author{D. A. Drabold}

\email{drabold@ohio.edu}

\affiliation{Department of Physics and Astronomy, Ohio University, Athens, OH 45701, USA }
\date{\today}%

\begin{abstract}
We employed \textit{ab-initio} molecular dynamics to directly simulate the effects of Ag alloying ($\sim5\%$ Ag concentration) on the phase change properties of \gst. The short range order is preserved, whereas a slight improvement in the chemical order is observed. A slight decrease in the fraction of tetrahedral Ge (sp$^{3}$ bonding) is reflected in the reduction of the optical band gap and in the increased dielectric constant. Simulations of the amorphous to crystalline phase change cycle revealed the fact that the crystallization speed in \gsta~ is no less than that in \gst. Moreover, the smaller density difference and the larger energy difference between the two phases of \gsta~ (compared to \gst) suggest a smaller residual stress in devices due to phase transition and improved thermal stability for \gsta. The potential viability of this material suggests the need for a wide exploration of alternative phase change memory materials.

\end{abstract}
%\pacs{61.43.-j 61.43.Bn 61.43.Fs}

\maketitle

%\section{Introduction}

The computational design of materials is still in its nascent stages, but is widely recognized to be one of the prime frontiers of materials science. The challenges are daunting for several reasons, among these: time and length scales drastically different in simulation compared to laboratory samples, the need for realistic interatomic interactions (nowadays largely based upon pseudopotentials and density functional theory) leads to tremendous demand for computational resources. In the case of the phase change memory materials, with compositions near \gst~(GST), there is clear evidence that current first principles simulations can accurately simulate phase changes on the time scales accessible to these codes\cite{elliott}. Other work suggests that key quantities like crystallization speed can be meaningfully inferred from such simulations\cite{elliott}. These materials are of great fundamental interest for their ultrafast phase changes and are the leading candidate to replace current non-volatile computer memory, a multi-billion dollar market.

In this paper, we explore new candidate phase change materials and show that a silver-doped variant may be superior to conventional GST. We elucidate the process of crystallization in atomistic detail and particularly note the role of the Ag in producing more stable and chemically ordered materials. Beside the specific prediction that the Ag alloy systems deserve careful exploration, we highlight the existence of a promising unexplored which strongly suggests that others probably exist as well, and deserve exploration.

We have implemented an \textit{ab-initio} molecular dynamic (AIMD) simulations to study the ultrafast crystallization of Ag-doped (alloyed)\gst. The AIMD calculations were performed using Vienna \textit{Ab-initio} Simulation Package (VASP)\cite{kresse, kresse2, kresse3} to generate models of \gst~ and \gsta~(AGST) with 108(24 Ge atoms, 24 Sb atoms and 60 Te atoms) and 114(24 Ge atoms, 24 Sb atoms, 60 Te atoms, and 6 Ag atoms) atoms, respectively. The calculations were performed by using the projector augmented-wave (PAW)\cite{paw,paw2} method to describe electron-ion interactions. The Perdew-Burke-Ernzerhof (PBE)\cite{pbe} exchange correlation functional was used throughout. Molecular-dynamics (MD) simulations were performed in a cubic supercell with a time step of 5.0 fs using periodic boundary conditions at constant volume for annealing, equilibrating and cooling, whereas, zero pressure conjugate gradient (CG) simulations were performed for relaxation. The final models were prepared by using the \textgravedbl Melt and Quench\textacutedbl~method\cite{dad} starting with a random configuration at 3000K. Densities of 6.046 gcm$^{-3}$ and 6.234 gcm$^{-3}$, respectively for GST and Ag-GST, were used during the process. After mixing the random configurations at 3000K for 20ps, each model was cooled to 1200K in 10ps and equilibrated for 60ps. A cooling rate of 12K/ps was adopted to obtain the amorphous models from the melt at 1200K to 300K and followed by equilibration at 300K for another 50ps. Finally the systems were fully relaxed to a local minimum  at 0 pressure. Each of these models was then equilibrated at 300K for 25ps and data was accumulated for the last 10ps and statistically averaged to study the structural properties. Three independent models were generated for both structures. Unless otherwise stated the results presented are the statistical average of the models.

\begin{figure}[th]
\par
\subfigure[]{\includegraphics[scale=0.5]{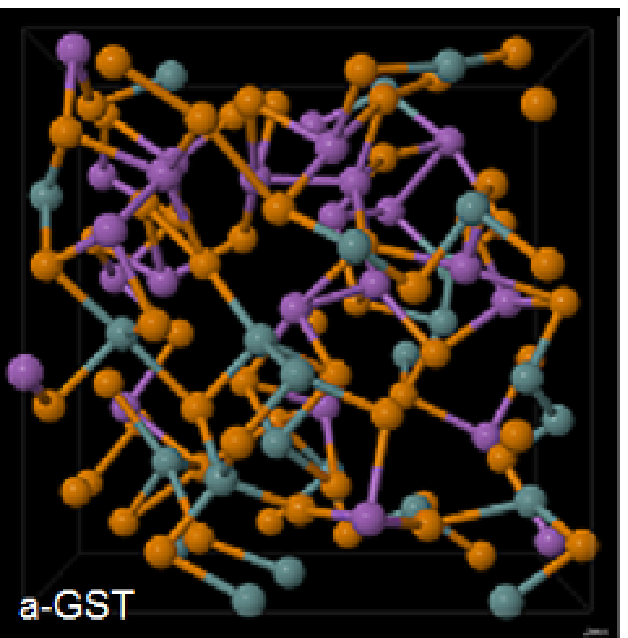}\label{gst}}\ %hfill %
\subfigure[]{\includegraphics[scale=0.5]{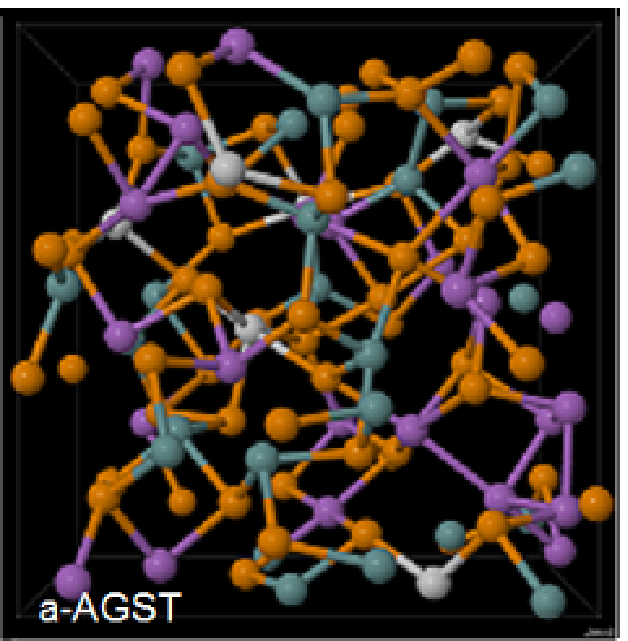}\label{gsta}}\hfill %
\subfigure[]{\includegraphics[scale=0.5]{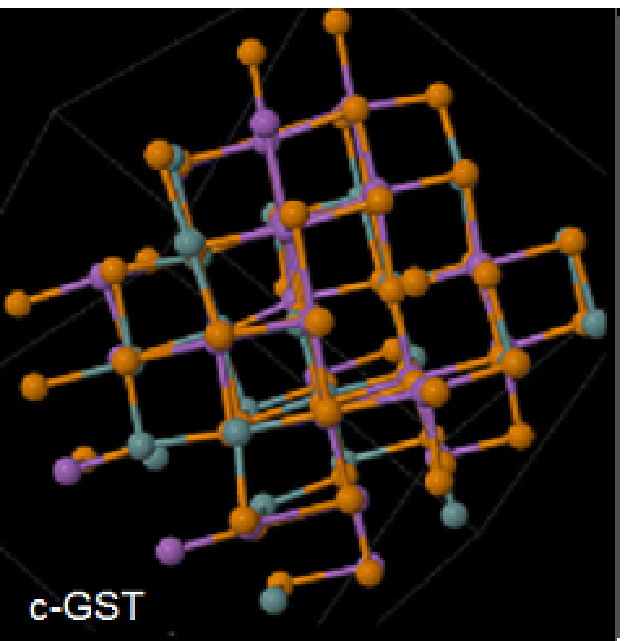}\label{xgst}}\ %hfill %
\subfigure[]{\includegraphics[scale=0.5]{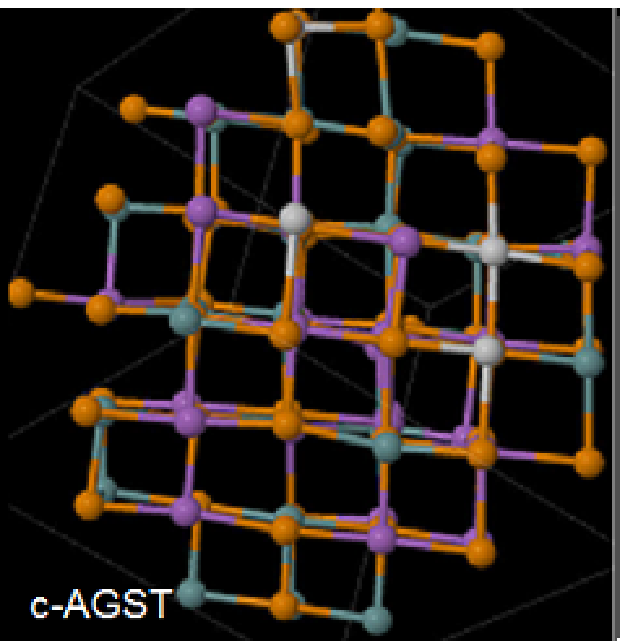}\label{xgsta}}\hfill %
\caption{(Color online) Simulated structures (relaxed) of a) a-\gst~, b) a-\gsta~, c)c-\gst~, and d) c-\gsta. (Model 3). Color code; Orange-Te, Green-Ge, Purple-Sb, and Gray-Ag.}
\label{snap}
\end{figure}

\begin{figure}
\includegraphics[width=0.45\textwidth]{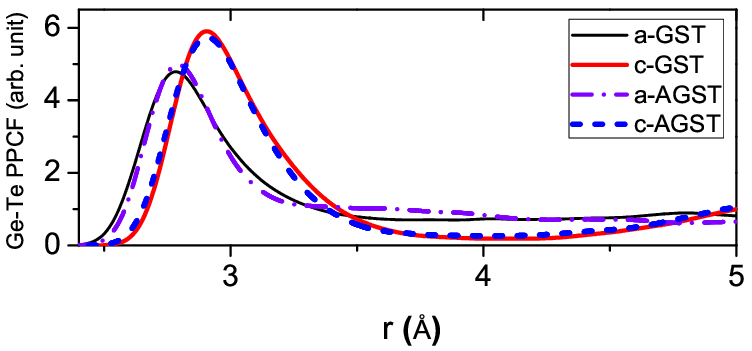}
\caption{\label{ppcf}(Color online) Ge-Te partial pair correlation functions (PPCFs) at 300K in both phases of \gst~ and \gsta. Only one model (Model 3) from each of the \gst~ and \gsta~ models is presented for the illustrations purpose.}
\end{figure}

\begin{figure}
\includegraphics[width=0.45\textwidth]{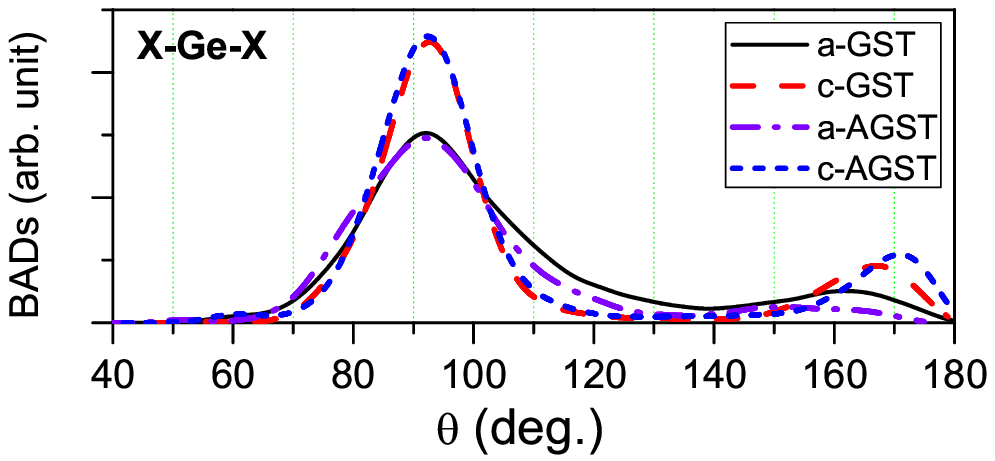}
\caption{\label{bad}(Color online) Ge-centered Bond angle distributions (BADs)(Model 3) in both phases of \gst~ and \gsta~ at 300K.}
\end{figure}

%\section{Result}

The zero-pressure structures of amorphous and crystalline GST and AGST are presented in Fig.\ref{snap}. Interestingly, the computation of the atomic densities shows a relatively small density change (4.61$\%$) between the two phases of AGST in contrast to a density change of 6.84 $\%$ in pure GST. This smaller volume (density) change in Ag-doped GST could result in reduced residual stress in PCM devices. We further computed the difference in the energies between the amorphous and the crystalline phases in GST and AGST. The energy difference of 80 meV/atom in AGST is about 20meV/atom more than that of pure GST. This larger energy difference might yield better thermal stability in Ag-doped GST, and could improve the data retention capability of PCM devices. To investigate the Ag-induced modification of GST network, we analyzed the local structure around Ge via partial pair correlation functions (PPCFs) and bond angle distributions (BAD). We inspected the Ge-Te PPCF (fig.\ref{ppcf}) and Ge-centered BAD (fig.\ref{bad}) because Ge atoms undergo a dramatic change upon phase transition \textit{i.e.} perfect octahedral configurations (p-bonding) in the crystalline phase and tetrahedral geometry(sp$^{3}$-bonding) in the amorphous phase\cite{kolobov}. In the amorphous phase, the Ge-Te bond length is found to be increased (by 0.02{\AA}) in Ag-doped GST. This Ag-induced change is also observed in the Ge-centered BAD as a suppression near 109$^{o}$ depicting a reduction in the fraction of tetrahedral Ge sites due to addition of Ag. This reduction caused the increase in the average Ge-Te bond length since Ge-Te bond lengths with tetrahedral geometry is smaller than Ge-Te with octahedral geometry \cite{raty,caravati2}. In contrast to the amorphous phase, we observed identical Ge-Te bond lengths between GST and AGST in the crystalline phases. This is consistent with the fact that all the tetrahedral Ge changes to an octahedral geometry during crystallization. On the other hand, Ag-induced change is negligible in the Sb-Te PPCF and Sb-centered BAD, in both phases. This is also supported by the fact that Sb always takes octahedral geometry in either of the phases. Beside Ge-Te and Sb-Te bonds pairs, we observed a significant fraction of Ge-Ge, Ge-Sb, Sb-Sb and Te-Te bond pairs as nearest neighbors especially in amorphous phase. These \textgravedbl wrong bonds\textacutedbl \cite{elliott}, amount to 25$\%$ in amorphous phases and fall to about 6$\%$ in the crystalline phases. From analysis of the local structure, we are also able to identify the interaction of the dopants (Ag atoms) in the host network. The Ag PPCF confirms that Ag is mainly bonded to Te rather than to Ge or Sb. This is also true in crystalline phase where Ag takes the vacancy sites(or similar sites as Ge/Sb).

The investigation of the electrical properties [via electronic density of states(EDOS)] in both phases of GST and AGST confirms no major differences in the EDOS, with p-like states of Te, Sb and Ge dominating both the valence and the conduction band and Ag contributing a d-like state about 4eV below the fermi level. The band gap is observed to decrease with the presence of Ag. Since the larger band gap in a-GST as compared to c-GST is due to the presence of sp$ ^{3} $-bonded Ge atoms\cite{akola}, the reduced band gap by doping can also be attributed to the reduction of the tetrahedral Ge atoms.

The utility of the PCMs stems from the contrast in optical properties between amorphous and crystalline phases. The imaginary part and the real part of the dielectric function confirm that the optical contrast is preserved in AGST. These results are in consistent with the results reported by Shportko \textit{et al.} \cite{shportko}. The estimation of the optical dielectric constant i.e. the lower energy-limit of the real part of the dielectric function ($\omega \rightarrow 0$) is presented in Table \ref{dielectric}. We observed a slightly higher dielectric constant in AGST as compared to GST and suspect that this is due to improved medium-range order (increase in the number of four-membered, near-square, rings\cite{elliott}) in AGST.

\begin{table}
\caption {Comparison of dielectric constant between the two phases of \gst~ and \gsta. (Model 3)}
\label{dielectric}%
\begin{ruledtabular}
\begin{tabular}{ccccccc}
Material&Amorphous&Crystalline&$\%$increase\\
\hline
\gst &25.9&53.0&105\\
\gst(Ref.\cite{shportko}) &16.0&33.3&108\\
\gsta&26.9&60.2&124\\
\end{tabular}
\end{ruledtabular}
\end{table}

\begin{figure}
\includegraphics[width=0.45\textwidth]{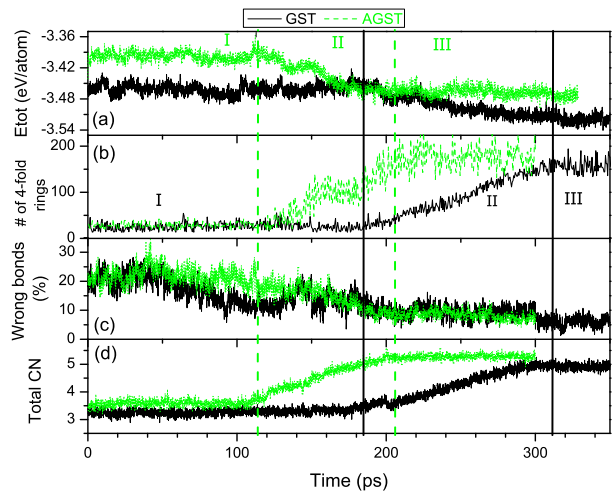}
\caption{\label{etot}(Color online) Comparison of the total energy (a), the number of four-fold rings (b), the number of wrong bonds (c), and the total coordination numbers (d) as functions of time in \gst~ (dark, black) and \gsta (light, green).(Model 3). The vertical lines separate the three periods.}
\end{figure}

\begin{figure*}
\includegraphics[width=0.6\textwidth]{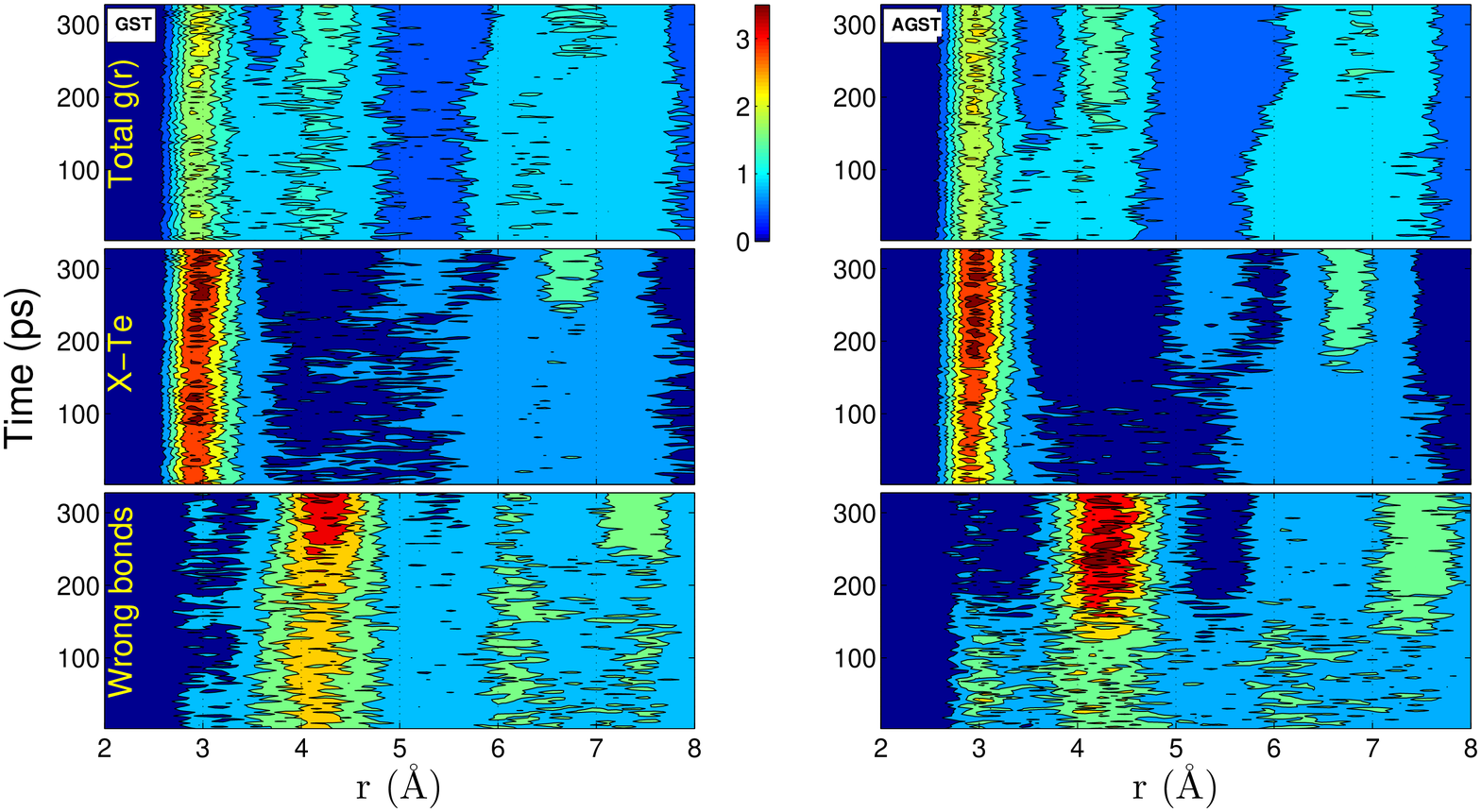}
\caption{\label{gr_contour}(Color online) Evolution of pair correlation functions in \gst~ (left) and \gsta~ (right) with time. (Model 3).}
\end{figure*}

\begin{figure}
\includegraphics[width=0.45\textwidth]{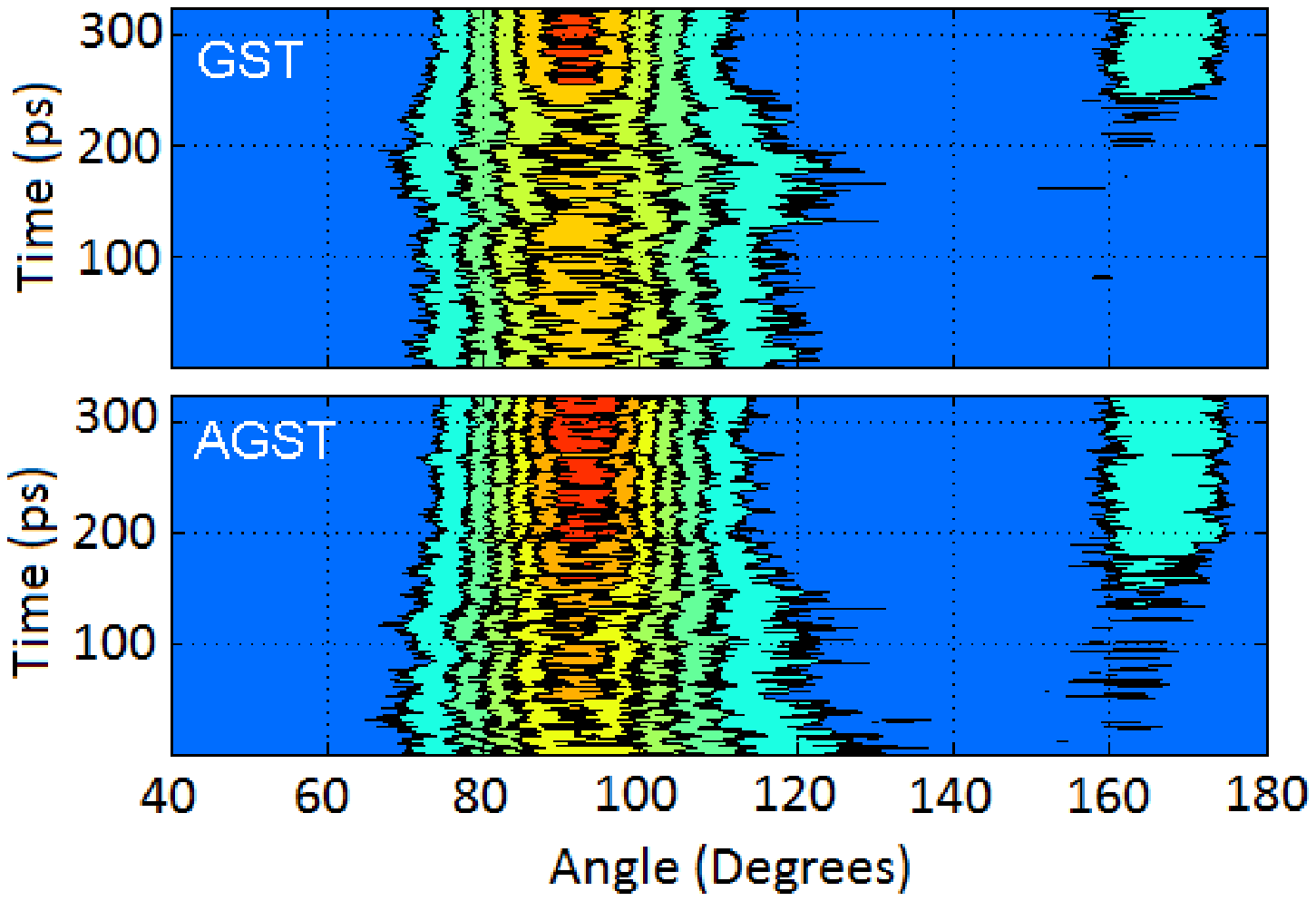}
\caption{\label{bad_contour}(Color online) Time evolution of Ge-centered bond angle distributions in \gst~ (left) and \gsta~ (right). (Model 3).}
\end{figure}

%section{dynamics}
The full potential of simulation is revealed in directly simulating phase transitions\cite{elliott,bin}. We annealed the a-GST and a-AGST models at 650K until each of the models crystallized. The process proceeds in three steps (I, II, and III), as explained by Lee \textit{et al.}\cite{lee}. Period I is termed the incubation period. Period II is the main time segment in which the process of crystallization occurs and the third Period (III) defines the completely crystallized state. To understand the crystallization process we observed the evolution of total energy of the system, the number of 4-member rings, seeds in the spirit of classical nucleation theory(CNT)\cite{lee}, the number of wrong bonds, and the total coordination numbers as a function of time and present in Fig.\ref{etot}(a-d). We observe almost no change in the total energy and the number of 4-member rings during the incubation period (I), however, we observe a significant decrease in the number of wrong bonds. Wrong bonds keep declining during the crystallization period (II) until the crystallization is complete. About 5 to 7 $\%$ of wrong bonds(mainly Ge-Sb and Sb-Sb bonds) still exist even after the crystallization is occurred. The total energy and the number of 4-member rings on the other hand are found to be correlated to each other, with the number of rings increasing monotonously during the crystallization period. We further computed the evolution of pair correlation functions and the Ge-centered bond angle distribution (for model 3) and present these findings in figures \ref{gr_contour} and \ref{bad_contour} respectively. The top panels of Fig.\ref{gr_contour} represent the total pair correlation functions (TPCF). The middle panel on the left represents  the X-Te (where X=Ge and Sb) pair correlation functions whereas that on the right represents X-Te (X=Ge, Sb, and Ag) PCFs. Finally, the bottom panels represent the correlation of wrong bonds (Y-Y, Y = Ge, Sb, or Ag). These figures clearly depict the evolution of medium to long range order (secondary peaks in PCFs) which is the signature of the crystalline structures. The prominent medium to long range order peaks start evolving during the Period II. The average peak positions at 2.98{\AA}, 5.3{\AA}, and 6.8{\AA} for X-Te and 4.2{\AA} and 7.4{\AA} for wrong bonds well represent the crystalline GST structure. Similarly, the Ge-centered BAD (Fig.\ref{bad_contour}) shows an evolution of narrow and prominent distribution around 90$^{o}$ and 180$^{o}$ during the Period II. The narrowing of the peak at 90$^{o}$ illustrates the conversion of the tetrahedral Ge (angular distribution at 109$^{o}$) into the octahedral Ge. The peak at around 180$^{o}$ becomes visible during the period II where the total coordination numbers(CN) reaches about 4.5, similar to the evolution of secondary peaks in PPCFs. CN on the other hand depicts a correlation with the total energy of the system \textit{i.e.} CN is almost constant during the incubation period (I), increases during the crystallization period (II), and becomes constant after the crystallization is established.

To investigate the effect of Ag on the crystallization we compare total energy, the number of four member rings and the coordination number (Fig.\ref{etot}). Since the crystallization of three different models of pure GST shows large fluctuations in the duration of Period I and II, especially Period I, the estimation of crystallization time shows significant uncertainty. The incubation periods (Period I) in three different pure GST models vary from 50 ps to 200ps whereas the crystallization periods (Period II) vary from 40ps to 150ps. These times in AGST are (80-110ps) for incubation periods and (70-110ps) for crystallization periods. To understand this we examined the local structures of the starting configuration of the three GST models. Unsurprisingly, we observe a clear distinction in the number of wrong bonds and four membered rings, i.e. the model with the short incubation and crystallization period has the least number of wrong bond and most number of four membered rings (more ordered in a plane). To compare the crystallization of GST and AGST we chose configurations with similar numbers of the wrong bonds. We observed a clear contrast in the duration of both the incubation period and the crystallization period in these two networks. Both of the periods were shorter in AGST than in GST. Total of these two periods in AGST measure about 200ps against about 315ps in pure GST, clearly suggesting a faster crystallization in Ag-doped GST.

%\section{Conclusion}
In conclusion, we have used AIMD simulations to study the effect of Ag doping in \gst~ and directly simulate ultrafast phase transitions. Medium range order is found to be improved with the addition of Ag in the form of increased number of four membered rings and decreased fraction of tetrahedral Ge. We were also able to simulate the process of amorphous to crystalline phase transition. The incubation and crystallization period were found to depend on the wrong bonds present in the amorphous phase. Moreover, our simulation revealed that the crystallization speed is increased by doping \gst~ with Ag. The larger energy/atom difference between amorphous and crystalline phases also suggests that \gsta~ is thermally more stable than \gst. On the other hand, smaller density difference in \gsta~ between the two phases as compared to \gst~ could well reduced the residual stress in the PCM devices. Furthermore, the increased optical contrast between the two phases as well as a potential increase in crystallization speed might lead to PCM devices with improved performance.

%\section{Acknowledgement}
We thank Prof. S. R. Elliott for many helpful conversations. This work was support by NSF DMR 09-03225 and NSF DMR 09-06825. This work was also supported in part by an allocation of computing time from the Ohio Supercomputer Center.

%\section{References}

\end{document}